\documentclass[aps,prb,amsmath,amssymb,twocolumn,superscriptaddress,showpacs,floatfix]{revtex4-1}

\usepackage{graphicx}
\usepackage{dcolumn}
\usepackage{bm}

\usepackage{amsmath}
\usepackage{amssymb}

\begin{document}

\title{Manipulation of Dzyaloshinskii-Moriya interaction in Co/Pt multilayers with strain}

\author{N.~S.~Gusev}
\affiliation{Institute for Physics of Microstructures RAS, Nizhny Novgorod, 603950, Russia}

\author{A.~V.~Sadovnikov}
\affiliation{Saratov State University, Saratov, 410012, Russia}

\author{S.~A.~Nikitov}
\affiliation{Saratov State University, Saratov, 410012, Russia}

\author{M.~V.~Sapozhnikov}
\affiliation{Institute for Physics of Microstructures RAS, Nizhny Novgorod, 603950, Russia}
\affiliation{Lobachevsky State University of Nizhny Novgorod, 603950 Nizhny Novgorod, Russia}

\author{O.~G.~Udalov}
\affiliation{Institute for Physics of Microstructures RAS, Nizhny Novgorod, 603950, Russia}
\affiliation{Lobachevsky State University of Nizhny Novgorod, 603950 Nizhny Novgorod, Russia}
\affiliation{Department of Physics and Astronomy, California State University Northridge, Northridge, CA 91330, USA}

\date{\today}

\pacs{75.50.Tt 75.75.Lf	75.30.Et 75.75.-c}

\begin{abstract}

Interfacial Dzyaloshinskii-Moriya interaction (DMI) is experimentally investigated in Pt/Co/Pt multilayer films under strain. A strong variation (from 0.1 to 0.8 mJ/m$^2$) of the DMI constant is demonstrated at $\pm 0.1\%$ in-plane uniaxial deformation of the films. The anisotropic strain induces strong DMI anisotropy. The DMI constant perpendicular to the strain direction changes sign while the constant along the strain direction does not. Estimates are made showing that DMI manipulation with an electric field can be realized in hybrid ferroelectric/ferromagnetic systems. So, the observed effect opens the way to manipulate the DMI and eventually skyrmions with a voltage via a strain-mediated magneto-electric coupling. 
\end{abstract}

\maketitle

Skyrmions in magnetic thin films  with perpendicular anistropy are non-trivial magnetic textures~[\onlinecite{RN33}] promising various applications such as memory and logics. Therefore, manipulating (creating, annihilating and moving) the skyrmions is an urge but still challenging quest of the modern spintronics~[\onlinecite{RN22,doi:10.1063/1.5048972,Finocchio_2016,RN26}]. So far, several approaches were used. Electrical current based techniques utilizing spin torque~[\onlinecite{Jiang283,RN24,RN25}] and spin-orbit torque~[\onlinecite{RN23,RN29}]  allow to control the skyrmions but require a high current density  and therefore, have low energy efficiency. A lot of groups work on electric field based approaches where the heat losses are minimized. One of the most actively studied approaches is based on voltage controlled magnetic anisotropy~[\onlinecite{Noh2018,993433,PhysRevB.97.024429,Nikitov2018,PhysRevB.97.224428}]. Since a skyrmion stability is defined by the competition of the magnetic anisotropy and the Dzyaloshinskii-Moriya interaction (DMI), tuning of one of these contributions opens the way to control the skyrmions. So far, people were focused on the variation of the magnetic anisotropy via a strain-mediated magneto-electric coupling~[\onlinecite{RN31}] or a charge-mediated magneto-electric effect~[\onlinecite{RevModPhys.89.025008}].

\begin{figure}[t]
\includegraphics[width=0.7\columnwidth, keepaspectratio]{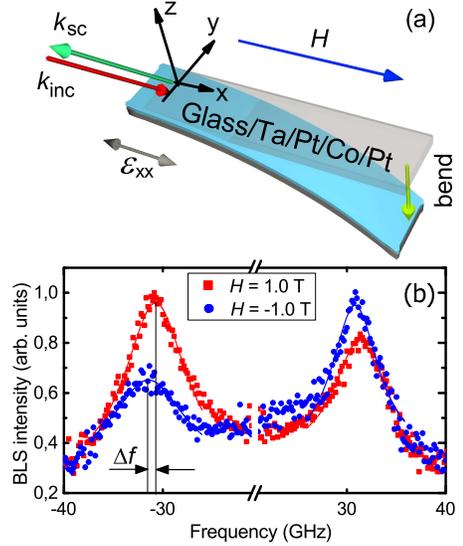}
\caption{(a) Experimental geometry. The sample (Glass/Ta/Pt/Co/Pt) is bent and has in-plane strain, $\varepsilon_{xx}$. BLS experiments are performed in the Damon-Eshbach geometry. The laser beam with the incident wavevector $\mathbf k_\mathrm{inc}$ (red arrow) laying in the ($y$,$z$)-plane irradiates the sample. The multilayers film scatters the light back into the direction $\mathbf k_\mathrm{sc}=-\mathbf k_\mathrm{inc}$ (green arrow). A magnetic field $H$ is applied perpendicular to the incidence plane. (b) Typical BLS spectrum of Glass/Ta/Pt/Co/Pt without a strain at $H=1$ T (squares) and $H=-1$ T (circles). Solid lines are Lorentzian fits. $\Delta f$ is the frequency shift between the Stokes and anti-Stokes peaks. \label{Fig:BLSexp}}%
\end{figure}

In the present work we experimentally demonstrate that the DMI can be also controlled with a strain. Previously, people studied strain dependence of the DMI in bulk crystals~[\onlinecite{Iwasa2015,RN20,Arita2015}]. Here, we  show that in heavy metal/ferromagnet (Co/Pt) multilayer structures the interfacial DMI coefficient can be tuned in a wide range by applying strain. The uniaxial strain modifies the average DMI constant and also introduces anisotropy to the DMI. Moreover, the DMI of different sign for different directions appears due to the uniaxial strain. 

Strains in magnetic films can be induced mechanically (via bending for example) or with an electric field in hybrid ferromagnetic (FM)/ferroelectric(FE) systems. In our work we use the mechanical mean to create the strain. At that, the deformations used in our experiments can be easily induced by an electric field in a ferroelectric (such as PMN-PT). This opens the way to control the DMI (and therefore skyrmions) in heavy metal(HM)/FM systems with voltage.

\begin{figure}[t]
\includegraphics[width=1.0\columnwidth]{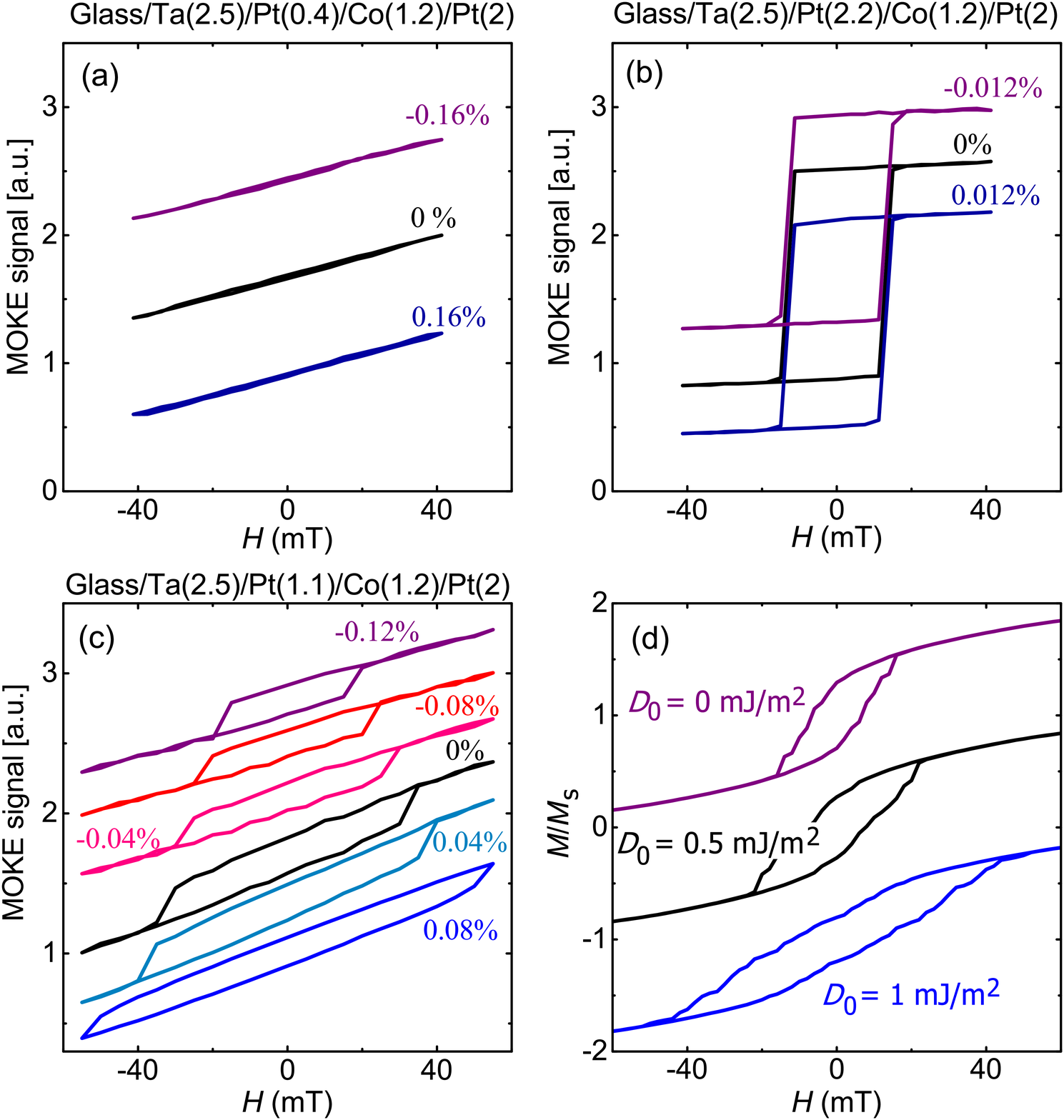}
\caption{Out-of-plane hysteresis loops for different strain $\varepsilon_\mathrm{xx}$ (shown nearby each curve) applied to the samples Glass/Ta(2)/Pt($d_\mathrm{Pt}$)/Co(1.2)/Pt(2). Panel (a) - $d_\mathrm{Pt}=0.4$ nm, (b) - $d_\mathrm{Pt}=1.1$ nm, (c) - $d_\mathrm{Pt}=2.2$ nm. (d) Micromagnetic simulation results for Co/Pt films. Top row is magnetization hysteresis loops for films with the different DMI: $D_0 = 0$ mJ/m$^2$, $D_0 = 0.5$ mJ/m$^2$, and $D_0 = 1$ mJ/m$^2$. The corresponding values of the hysteresis widths are 34, 44 and 86 mT. The loops are shifted with respect to each other for clarity. \label{Fig:MSptThickness}}%
\end{figure}

Note that voltage based tuning of the DMI due to a charge accumulation was demonstrated in a Pt/Co/TaO multilayers in Ref.~[\onlinecite{Bea2018}]. The DMI in this system appears at the insulator/FM boundary rather than at the HM/FM interface. Therefore, the DMI in this system is much weaker (of order of 0.1 mJ/m$^2$) than in HM/FM multilayers (of order of 1 mJ/m$^2$). This restricts using of insulator/FM systems in skyrmionics. Voltage-induced variation of the DMI due to the charge accumulation is challenging in HM/FM multilayers since the electric field is screened in a very thin interfacial layer. In contrast, the strain-based approach proposed in the present work can be applied to metallic system giving a promising opportunity to control the skyrmions.

In the present work, a series of samples Glass/Ta(2.5 nm)/Pt($d_\mathrm{Pt}$)/Co(1.2 nm)/Pt(2 nm) were fabricated using DC magnetron sputtering. The thickness of the bottom Pt layer ($d_\mathrm{Pt}$) varies from 0.4 to 2.2 nm. Fabricating samples with different Pt thickness allows to find the one which is the most sensitive to a strain. In our samples the Co film are surrounded by two Pt layers. One can expect that DMI cancels in this case. However, well known that the nonzero DMI is observed in such symmetric Pt/Co/Pt systems ~[\onlinecite{PhysRevB.99.014433}]. This is because Pt/Co and Co/Pt interfaces actually are not identical, since the bottom Pt layer grows on the Ta buffer, while the upper Pt layer grows on Co. Moreover the DMI strongly depends on Pt thickness ~[\onlinecite{PhysRevLett.118.147201}] which also makes the contributions of the upper and bottom interfaces different. 

Magnetic hysteresis loops of the samples were measured at different in-plane uniaxial strain using a magneto-optical Kerr effect (MOKE) in polar geometry. A sample was placed inside the specially designed holder (see Fig.~\ref{Fig:BLSexp}(a)). One edge of the sample was fixed in the holder, the opposite edge was bent by a screw inducing a uniaxial strain. The strain is elastic and does not produce a damage to the samples (see Supplementary materials). The shift of the sample free edge caused a strain of the magnetic film in the vicinity of the fixed side, where the laser beam irradiates the film. Introducing the x-axis connecting fixed and free edges (Fig.~\ref{Fig:BLSexp}(a)) one can estimate the x-component of the strain as  $\varepsilon_{xx}=3d\Delta z/(2L^2)$ [\onlinecite{LL_elast}], where $d$ and $L$ are the thickness and length of the sample (glass plate), correspondingly, and $\Delta z$ is the shift of the plate free end.  The in-plane deformation was also checked using a strain gauge.

Figure~\ref{Fig:MSptThickness} shows the magnetization curves of the Co/Pt samples for different $d_\mathrm{Pt}$. Each panel in Fig.~\ref{Fig:MSptThickness} demonstrates several hysteresis loops corresponding to different strain, $\varepsilon_{xx}$.  The panel (a) shows the case of small Pt thickness, in which the structure has an in-plane anisotropy and is not sensitive to the applied strain. The sample with the thick Pt layer (panel (b)) has a rectangular magnetization curve and no magnetostriction. The strain influences the properties of the film only when the Pt layer is close to the critical thickness at which transition between in-plane and out-of-plane anisotropy occurs. This case is shown in panel (c). The curves in this plot consist of a linear slope and a hysteresis loop. Black line in the panel represents the unstrained film. Compressive strain increases the hysteresis loop width while tensile strain reduces it. Two additional samples were also studied with the thickness of Pt layer in the range between 1.1 and 2 nm. They have a hysteresis loop similar to the sample with $d_\mathrm{Pt}=1.1$ nm. They also demonstrate the dependence of the hysteresis loop on the strain.

The DMI in the samples was studied by the Brillouin light scattering (BLS) in the Damon-Eshbach geometry~[\onlinecite{PhysRevLett.114.047201}] under application of strain in the similar way as described in the previous section (see Fig.~\ref{Fig:BLSexp}(a)). A magnetic field was applied either along the deformation direction or perpendicular to it allowing us to measure the DMI constants along $x$ ($D_x$) and along $y$ ($D_y$) directions. Typical BLS spectrum is presented in Fig.~\ref{Fig:BLSexp}(b). Solid lines show the Lorentzian fit demonstrating the shift of the Stokes and anti-Stokes peaks denoted as $\Delta f$. 

\begin{figure}[t]
\includegraphics[width=0.9\columnwidth, keepaspectratio]{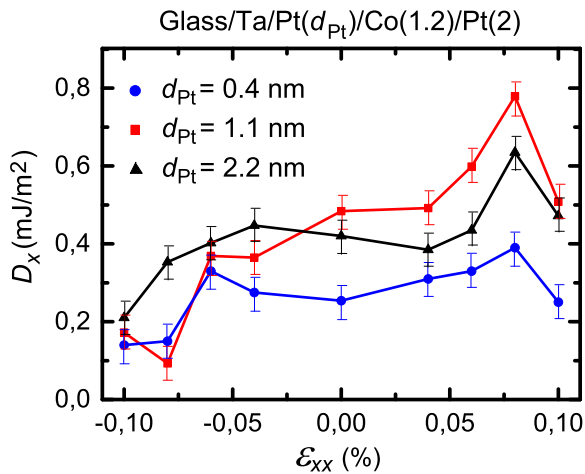}
\caption{The DMI constant measured along the $x$-direction ($D_x$) as a function of applied strain ($\varepsilon_{xx}$). See the linear least-square fits in the Supplementary materials. \label{Fig:DMIallSamples}}%
\end{figure}

Following the standard approach (see Supplementary materials) we estimated the DMI constant as [\onlinecite{PhysRevLett.114.047201, PhysRevB.88.184404}]
\begin{equation}\label{Eq:DMI}
D_i=2 M_\mathrm s\Delta f/(\pi \gamma k_i),
\end{equation}
where $M_\mathrm s$ is the saturation magnetization, $\Delta f$ is the difference between the Stokes and anti-Stokes frequencies, and $k_i$ is the momentum along the $i$-direction (in our case $i=x$ or $y$), and $\gamma$ is the gyromagnetic ratio. The value of $M_\mathrm s$ used in our estimations is $1.1\cdot 10^6$ A/m which is typical for Co/Pt films~[\onlinecite{Ding2013,Karis2012}].  

The DMI constant along the $x$-direction for the three samples with the Pt thickness varying from 0.4 nm to 2.2 nm is shown in Fig.~\ref{Fig:DMIallSamples} as a function of strain, $\varepsilon_{xx}$. The samples with $d_\mathrm{Pt}=0.4$ and 2.2 nm show weak variation of the DMI. The sample with $d_\mathrm{Pt}=1.1$ nm demonstrates rather strong change of the DMI constant $D_x$ from 0.1 to 0.8 mJ/m$^2$ which is 8 times variation. Note that $D_x=0.8$ mJ/m$^2$ is the DMI constant high enough for stabilization of skyrmions in Co/Pt systems~[\onlinecite
{RN22}] while 0.1 mJ/m$^2$ is too low for skyrmion formation. Therefore, one can effectively control the skyrmions using the strain induced DMI modulation. 

\begin{figure}[t]
\includegraphics[width=0.9\columnwidth, keepaspectratio]{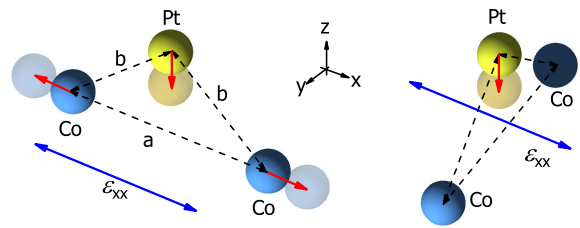}
\caption{Displacement of Co and Pt ions due to xx tensile strain $\varepsilon_{xx}$. $a$ is the distance between the Co ions. $b$ is the distance between the Pt and Co ions. Left panel shows the ion triangle oriented along the x-axis. Right panel shows the triangle oriented perpendicular to the strain axis. \label{Fig:DMImodel}}%
\end{figure}

The microscopic reason for the DMI strain dependence can be understood on the base the theoretical model by Fert and Levy~[\onlinecite{PhysRevLett.44.1538}]. According to this model the DMI is mediated by conducting electrons hopping between magnetic ions through heavy metal ions. Since the interaction appears due to the conduction electrons, it has oscillating character and is described by the expression
\begin{equation}\label{Eq:LevyFert}
W_\mathrm{DMI}\sim \sin(k_\mathrm{F}(a+2b)+\pi Z_\mathrm d/10)\sin(2\theta)/(ab^2),
\end{equation}
where $k_\mathrm F$ is the Fermi momentum, $a$ is the distance between magnetic (Co) ions (see Fig.~\ref{Fig:DMImodel}), $b$ is the distance between magnetic and heavy metal (Pt) ions, $Z_\mathrm d$ is the number of d-electrons, and $\theta$ is the angle made by vectors connecting heavy metal ion and two magnetic ions.

The in-plane strain  produced by bending changes the distances $a$ and $b$. For example, the tensile strain along the x-axis increases $a$ but decreases the height of Pt ion (see left panel in Fig.~\ref{Fig:DMImodel}). The height reduces according to the Poisson law. This modifies the DMI constant. Eq.~(\ref{Eq:LevyFert}) gives non-monotonic behaviour of the DMI constant as a function of distances. This probably is the reason for the non monotonic behaviour of the DMI constant at a high strain.

Note however, that the proposed consideration does not explain the dependence of the DMI strain variation on the Pt layer thickness. At first the model includes only one neighbouring Pt layer, while all other layers may contribute. Another factor is that $d_{\mathrm{Pt}}$ influences the lattice constant $a$ in the Pt layer closest to the Co film.

\begin{figure}[t]
\includegraphics[width=0.9\columnwidth, keepaspectratio]{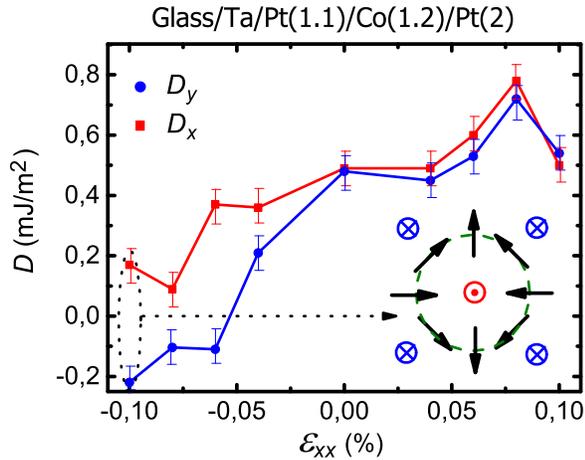}
\caption{The DMI constant measured along the x- and y-directions ($D_{x,y}$) as a function of applied strain ($\varepsilon_{xx}$) for the sample with $d_{\mathrm{Pt}}=1.1$ nm. See the linear least-square fits in the Supplementary materials. The inset shows the skyrmion with an antivortex domain wall. It may appear due to the anisotropic DMI with different sign along different directions~[\onlinecite{PhysRevB.95.214422}].\label{Fig:DMIanisotropy}}%
\end{figure}

Since the strain induced in our samples is anisotropic, one can expect that the DMI is also anisotropic. This is demonstrated in Fig.~\ref{Fig:DMIanisotropy} where the behaviour of the DMI coefficient for two different directions is shown for the sample with $d_\mathrm{Pt}=1.1$ nm. The uniaxial deformation changes the DMI coefficient for both directions. For the tensile strain $D_x\approx D_y$ but for the compressive strain there is a strong anisotropy of the DMI coefficient $D_x\ne D_y$.

The DMI anisotropy can be also understood from the crude consideration on the base of Eq.~(\ref{Eq:LevyFert}). When the deformation is applied along x-axis the DMI constant along this direction is modified due to variation of both $a$ and $b$ (see left panel in Fig.~\ref{Fig:DMImodel}). At that the DMI constant in the y-direction is defined by ion triangles along y-axis. These triangles are modified in a different way (see right panel in Fig.~\ref{Fig:DMImodel}). The distance between magnetic ions $a$ is not changed, while the height of the Pt ion reduces. So, variation of the DMI constant in this direction is different.

What is even more interesting is that at strong compressive strain the $y$-component of the DMI 
changes the sign while the $x$-component does not.  In Ref.~[\onlinecite{PhysRevB.95.214422}] authors simulate the magnetic skyrmions in the situation with different sign of the DMI along different directions. They show that the skyrmion with an anti-vortex domain wall (see inset in Fig.~\ref{Fig:DMIanisotropy}) can be realized in this case. So, the strained Co/Pt films can be a good candidates for studying such ``antivortex'' skyrmions.

Usually, the interface induced DMI in the thin film is described by the expression $-D(\mathbf m\cdot [[\mathbf z\times \nabla]\times m])$, where $D$ is the DMI constant, $\mathbf m$ is the normalized magnetization vector and $\mathbf z$ is the interface normal. This expression describes the system isotropic in the film plane. In our study we use uniaxial strain inducing the anisotropic DMI. The interaction energy $W_\mathrm{DMI}$ can be described by the expression
\begin{equation}\label{Eq:DMIinter}
\begin{split}
W_\mathrm{DMI}\!=&D_x\!\!\left(\!\!m_x\frac{\partial m_z}{\partial x}-m_z\frac{\partial m_x}{\partial x}\!\!\right)\!
\!+\!\!D_y\!\!\left(\!\!m_y\frac{\partial m_z}{\partial y}-m_z\frac{\partial m_y}{\partial y}\!\!\right)\!.
\end{split}
\end{equation}
In the linear approximation the constants $D_{x,y}$ can be expressed via strain as follows
\begin{equation}\label{Eq:DMIvector}
D_{x,y} = D_{0x,y}+D_1(\varepsilon_{xx}+\varepsilon_{yy})\pm D_{\mathrm{an}}(\varepsilon_{xx}-\varepsilon_{yy}),
\end{equation}
where the tensor $\varepsilon$ is the strain in the film, the sign ``+'' (``-'') is for $D_x$ ($D_y$). The first term describes the anisotropic DMI in the unstrained film, the second term shows the influence of the isotropic strain and the third contribution represents the effect of the anisotropic deformation.

Using a linear least-squares fit (see Supplementary materials) of our data we get the constants $D_{0x}$, $D_1$, and $D_\mathrm{an}$ for our samples. The obtained results are summarized in the Table~\ref{Table:DMI}. First three lines are for samples shown in Fig.~\ref{Fig:DMIallSamples}. Two bottom lines are for two additional samples mentioned above. The second column indicate the anisotropy type in each sample. Two additional samples studied here have the ``mixed'' type of magnetic hysteresis loop similar to the sample with $d_\mathrm{Pt}=1.1$ nm.
While the uncertainty of the data is quite hight, all the samples have non zero sensitivity to the strain (see $D_1+D_\mathrm{an}$). The samples with the mixed anisotropy type ($d_\mathrm{Pt}=1.1$ nm and $d_\mathrm{Pt}=1.9$ nm) have the highest average sensitivity. The mixed type of the anisotropy and high average $D_1+D_\mathrm{an}$ appear in the samples with intermediate Pt thickness. The samples with thin small ($d_\mathrm{Pt}=0.4$ nm) and high ($d_\mathrm{Pt}=2.2$ nm) Pt thickness have lower strain sensitivity. The films with the mixed anisotropy type demonstrate strong DMI anisotropy also (see $D_\mathrm{an}$).

\begin{table}
\caption{DMI interaction constants for different samples. The first three lines show the data for the sample in Fig.~\ref{Fig:DMIallSamples} and \ref{Fig:DMIanisotropy}. The last two lines show the data for two additional samples. The constants $D_{0x}$, $D_{0y}$ are measured in mJ/m$^2$, $D_1$, $D_\mathrm{an}$ are measured in mJ/(m$^2$(\%)). The samples thickness is defined with the precision of 20\%.}
\begin{tabular}{|c|c|c|c|c|c|c|}
\hline \label{Table:DMI}
$d_\mathrm{Pt}$, nm & Anis. & $D_{0x}$ & $D_{0y}$ & $D_{1}$ & $D_\mathrm{an}$ & $D_{1}+D_\mathrm{an}$ \\ 
\hline 
0.4 & in-plane & 0.27$\pm$0.03 & ~~~-~~~ & ~~~-~~~ & ~~~-~~~ & 0.7$\pm$0.6
 \\ 
\hline 
1.1 & mixed & 0.43 & 0.3 & 3.4 & -0.9 & 2.5 \\ 
&&$\pm$0.08&$\pm$0.1&$\pm$1.1&$\pm$0.6& $\pm$1.1\\
\hline 
1.9 & mixed & 0.4 & 0.2 & 3.2 & -1 & 2.5 \\
&&$\pm$0.05 &$\pm$0.1&$\pm$0.7&$\pm$0.8& $\pm$ 1.4\\ 
\hline 
2 & mixed & 0.42 & 0.4 & 2.1 & -0.5 & 1.6 \\
&&$\pm$0.02& $\pm$0.03&$\pm$1&$\pm$0.3& $\pm$0.8\\ 
\hline
2.2 & perp. & 0.42$\pm$0.03 & - & - & - & 1.1$\pm$0.8 \\ 
\hline  
\end{tabular}
\end{table} 

The strain induced in our films due to the bending of the samples is of order of $0.1\%$. Such a value can be easily achieved in ferroelectric crystals under application of voltage. For example, in Pb(Mg$_{1/3}$Nb$_{2/3}$)O$_3$–PbTiO$_3$ (PMN-PT) crystal the voltage induced strain reaches 0.3$\%$~[\onlinecite{Carman}] which is even higher than what we use in our experiments. So, one can control DMI with voltage in ferroelectric/(Co/Pt) systems. Assuming linear dependence of the DMI coefficient on $\varepsilon$ one can expect modulation of the DMI constant from -0.8 to 1.8 mJ/m$^2$ in the electric field range of about $\pm 600$ V/mm in PMN-PT/Ta/Pt/Co/Pt system. Note that for certain cut of PMN-PT crystal the induced strain is highly anisotropic. So, the voltage controlled DMI anisotropy can be realized.

To understand the correlation between the strain-induced DMI variations in Fig.~\ref{Fig:DMIallSamples} and the magnetization curves transformations in Fig.~\ref{Fig:MSptThickness}c, we carried out micromagnetic simulations using the OOMMF code~[\onlinecite{OOMMF}]. The results are shown in Fig.~\ref{Fig:MSptThickness}d. In the simulations we assumed the isotropic DMI varying with the strain similarly to what we observed in our BLS experiments ($D_0 = 0$, $0.5$, and $1$ mJ/m$^2$). The saturation magnetization $M_\mathrm s = 1.1\cdot 10^6$ A/m and the exchange stiffness $A = 2\cdot 10^{11}$ J/m~[\onlinecite{Karis2012}] were uniform across the film.  The magnitude of the perpendicular uniaxial anisotropy varies across the sample between $K_\mathrm{min} = 6.3\cdot 10^5$ J/m$^3$ and $K_\mathrm{max} = 8.3\cdot 10^5$ J/m$^3$.  These values are near the critical anisotropy  $K=\mu_0 M_\mathrm s^2/2 = 7.6\cdot 10^5$ J/m$^3$ corresponding to the easy plain - easy axis transition. The parameters used are in agreement with what we obtained from fitting of BLS data. The BLS data  confirms also that the anisotropy varies weakly with strain (see Supplementary materials).

Increasing the DMI reduces the domain wall energy and increases the magnetic field at which domains disappear (the hysteresis loop width). This is in agreement with our experimental observations (Fig.~\ref{Fig:MSptThickness}). So, we conclude that the magnetization loops variations observed for the film with the intermediate anisotropy are in good agreement with our BLS data.

In summary, we performed BLS and MOKE studies of strained Pt/Co/Pt films. We demonstrated that the strain strongly influences the DMI in the system. Application of $\pm 0.1\%$ in-plane deformations varies the DMI constant from 0.1 to 0.8 mJ/m$^2$. Moreover, strong DMI anisotropy appears under compressive strain. The DMI constant perpendicular to the strain direction changes sign while the constant along the strain direction  does not. The magnetic film with the DMI of opposite sign along  directions perpendicular to each other is suitable for realization of skyrmions with an antivortex domain wall. The strain used in the present work is less than what can be achieved in a hybrid FE/FM system. This opens the way to manipulate the DMI and eventually the skyrmions with a voltage via the strain-mediated magneto-electric coupling. 

This research was supported by the Russian Science Foundation (Grant  18-72-10026).

\bibliography{DMI}

\begin{thebibliography}{32}%
\makeatletter
\providecommand \@ifxundefined [1]{%
 \@ifx{#1\undefined}
}%
\providecommand \@ifnum [1]{%
 \ifnum #1\expandafter \@firstoftwo
 \else \expandafter \@secondoftwo
 \fi
}%
\providecommand \@ifx [1]{%
 \ifx #1\expandafter \@firstoftwo
 \else \expandafter \@secondoftwo
 \fi
}%
\providecommand \natexlab [1]{#1}%
\providecommand \enquote  [1]{``#1''}%
\providecommand \bibnamefont  [1]{#1}%
\providecommand \bibfnamefont [1]{#1}%
\providecommand \citenamefont [1]{#1}%
\providecommand \href@noop [0]{\@secondoftwo}%
\providecommand \href [0]{\begingroup \@sanitize@url \@href}%
\providecommand \@href[1]{\@@startlink{#1}\@@href}%
\providecommand \@@href[1]{\endgroup#1\@@endlink}%
\providecommand \@sanitize@url [0]{\catcode `\\12\catcode `\$12\catcode
  `\&12\catcode `\#12\catcode `\^12\catcode `\_12\catcode `\%12\relax}%
\providecommand \@@startlink[1]{}%
\providecommand \@@endlink[0]{}%
\providecommand \url  [0]{\begingroup\@sanitize@url \@url }%
\providecommand \@url [1]{\endgroup\@href {#1}{\urlprefix }}%
\providecommand \urlprefix  [0]{URL }%
\providecommand \Eprint [0]{\href }%
\providecommand \doibase [0]{http://dx.doi.org/}%
\providecommand \selectlanguage [0]{\@gobble}%
\providecommand \bibinfo  [0]{\@secondoftwo}%
\providecommand \bibfield  [0]{\@secondoftwo}%
\providecommand \translation [1]{[#1]}%
\providecommand \BibitemOpen [0]{}%
\providecommand \bibitemStop [0]{}%
\providecommand \bibitemNoStop [0]{.\EOS\space}%
\providecommand \EOS [0]{\spacefactor3000\relax}%
\providecommand \BibitemShut  [1]{\csname bibitem#1\endcsname}%
\let\auto@bib@innerbib\@empty
\bibitem [{\citenamefont {Nagaosa}\ and\ \citenamefont {Tokura}(2013)}]{RN33}%
  \BibitemOpen
  \bibfield  {author} {\bibinfo {author} {\bibfnamefont {N.}~\bibnamefont
  {Nagaosa}}\ and\ \bibinfo {author} {\bibfnamefont {Y.}~\bibnamefont
  {Tokura}},\ }\href@noop {} {\bibfield  {journal} {\bibinfo  {journal} {Nature
  Nanotechnology}\ }\textbf {\bibinfo {volume} {8}},\ \bibinfo {pages} {899}
  (\bibinfo {year} {2013})}\BibitemShut {NoStop}%
\bibitem [{\citenamefont {Fert}\ \emph {et~al.}(2017)\citenamefont {Fert},
  \citenamefont {Reyren},\ and\ \citenamefont {Cros}}]{RN22}%
  \BibitemOpen
  \bibfield  {author} {\bibinfo {author} {\bibfnamefont {A.}~\bibnamefont
  {Fert}}, \bibinfo {author} {\bibfnamefont {N.}~\bibnamefont {Reyren}}, \ and\
  \bibinfo {author} {\bibfnamefont {V.}~\bibnamefont {Cros}},\ }\href@noop {}
  {\bibfield  {journal} {\bibinfo  {journal} {Nature Reviews Materials}\
  }\textbf {\bibinfo {volume} {2}},\ \bibinfo {pages} {17031} (\bibinfo {year}
  {2017})}\BibitemShut {NoStop}%
\bibitem [{\citenamefont {Everschor-Sitte}\ \emph {et~al.}(2018)\citenamefont
  {Everschor-Sitte}, \citenamefont {Masell}, \citenamefont {Reeve},\ and\
  \citenamefont {Klaui}}]{doi:10.1063/1.5048972}%
  \BibitemOpen
  \bibfield  {author} {\bibinfo {author} {\bibfnamefont {K.}~\bibnamefont
  {Everschor-Sitte}}, \bibinfo {author} {\bibfnamefont {J.}~\bibnamefont
  {Masell}}, \bibinfo {author} {\bibfnamefont {R.~M.}\ \bibnamefont {Reeve}}, \
  and\ \bibinfo {author} {\bibfnamefont {M.}~\bibnamefont {Klaui}},\
  }\href@noop {} {\bibfield  {journal} {\bibinfo  {journal} {Journal of Applied
  Physics}\ }\textbf {\bibinfo {volume} {124}},\ \bibinfo {pages} {240901}
  (\bibinfo {year} {2018})}\BibitemShut {NoStop}%
\bibitem [{\citenamefont {Finocchio}\ \emph {et~al.}(2016)\citenamefont
  {Finocchio}, \citenamefont {Büttner}, \citenamefont {Tomasello},
  \citenamefont {Carpentieri},\ and\ \citenamefont {Klaui}}]{Finocchio_2016}%
  \BibitemOpen
  \bibfield  {author} {\bibinfo {author} {\bibfnamefont {G.}~\bibnamefont
  {Finocchio}}, \bibinfo {author} {\bibfnamefont {F.}~\bibnamefont {Büttner}},
  \bibinfo {author} {\bibfnamefont {R.}~\bibnamefont {Tomasello}}, \bibinfo
  {author} {\bibfnamefont {M.}~\bibnamefont {Carpentieri}}, \ and\ \bibinfo
  {author} {\bibfnamefont {M.}~\bibnamefont {Klaui}},\ }\href@noop {}
  {\bibfield  {journal} {\bibinfo  {journal} {Journal of Physics D: Applied
  Physics}\ }\textbf {\bibinfo {volume} {49}},\ \bibinfo {pages} {423001}
  (\bibinfo {year} {2016})}\BibitemShut {NoStop}%
\bibitem [{\citenamefont {Soumyanarayanan}\ \emph {et~al.}(2017)\citenamefont
  {Soumyanarayanan}, \citenamefont {Raju}, \citenamefont {Gonzalez~Oyarce},
  \citenamefont {Tan}, \citenamefont {Im}, \citenamefont {Petrovic},
  \citenamefont {Ho}, \citenamefont {Khoo}, \citenamefont {Tran}, \citenamefont
  {Gan}, \citenamefont {Ernult},\ and\ \citenamefont {Panagopoulos}}]{RN26}%
  \BibitemOpen
  \bibfield  {author} {\bibinfo {author} {\bibfnamefont {A.}~\bibnamefont
  {Soumyanarayanan}}, \bibinfo {author} {\bibfnamefont {M.}~\bibnamefont
  {Raju}}, \bibinfo {author} {\bibfnamefont {A.~L.}\ \bibnamefont
  {Gonzalez~Oyarce}}, \bibinfo {author} {\bibfnamefont {A.~K.~C.}\ \bibnamefont
  {Tan}}, \bibinfo {author} {\bibfnamefont {M.-Y.}\ \bibnamefont {Im}},
  \bibinfo {author} {\bibfnamefont {A.~P.}\ \bibnamefont {Petrovic}}, \bibinfo
  {author} {\bibfnamefont {P.}~\bibnamefont {Ho}}, \bibinfo {author}
  {\bibfnamefont {K.~H.}\ \bibnamefont {Khoo}}, \bibinfo {author}
  {\bibfnamefont {M.}~\bibnamefont {Tran}}, \bibinfo {author} {\bibfnamefont
  {C.~K.}\ \bibnamefont {Gan}}, \bibinfo {author} {\bibfnamefont
  {F.}~\bibnamefont {Ernult}}, \ and\ \bibinfo {author} {\bibfnamefont
  {C.}~\bibnamefont {Panagopoulos}},\ }\href@noop {} {\bibfield  {journal}
  {\bibinfo  {journal} {Nature Materials}\ }\textbf {\bibinfo {volume} {16}},\
  \bibinfo {pages} {898} (\bibinfo {year} {2017})}\BibitemShut {NoStop}%
\bibitem [{\citenamefont {Jiang}\ \emph {et~al.}(2015)\citenamefont {Jiang},
  \citenamefont {Upadhyaya}, \citenamefont {Zhang}, \citenamefont {Yu},
  \citenamefont {Jungfleisch}, \citenamefont {Fradin}, \citenamefont {Pearson},
  \citenamefont {Tserkovnyak}, \citenamefont {Wang}, \citenamefont {Heinonen},
  \citenamefont {te~Velthuis},\ and\ \citenamefont {Hoffmann}}]{Jiang283}%
  \BibitemOpen
  \bibfield  {author} {\bibinfo {author} {\bibfnamefont {W.}~\bibnamefont
  {Jiang}}, \bibinfo {author} {\bibfnamefont {P.}~\bibnamefont {Upadhyaya}},
  \bibinfo {author} {\bibfnamefont {W.}~\bibnamefont {Zhang}}, \bibinfo
  {author} {\bibfnamefont {G.}~\bibnamefont {Yu}}, \bibinfo {author}
  {\bibfnamefont {M.~B.}\ \bibnamefont {Jungfleisch}}, \bibinfo {author}
  {\bibfnamefont {F.~Y.}\ \bibnamefont {Fradin}}, \bibinfo {author}
  {\bibfnamefont {J.~E.}\ \bibnamefont {Pearson}}, \bibinfo {author}
  {\bibfnamefont {Y.}~\bibnamefont {Tserkovnyak}}, \bibinfo {author}
  {\bibfnamefont {K.~L.}\ \bibnamefont {Wang}}, \bibinfo {author}
  {\bibfnamefont {O.}~\bibnamefont {Heinonen}}, \bibinfo {author}
  {\bibfnamefont {S.~G.~E.}\ \bibnamefont {te~Velthuis}}, \ and\ \bibinfo
  {author} {\bibfnamefont {A.}~\bibnamefont {Hoffmann}},\ }\href@noop {}
  {\bibfield  {journal} {\bibinfo  {journal} {Science}\ }\textbf {\bibinfo
  {volume} {349}},\ \bibinfo {pages} {283} (\bibinfo {year}
  {2015})}\BibitemShut {NoStop}%
\bibitem [{\citenamefont {Hrabec}\ \emph {et~al.}(2017)\citenamefont {Hrabec},
  \citenamefont {Sampaio}, \citenamefont {Belmeguenai}, \citenamefont {Gross},
  \citenamefont {Weil}, \citenamefont {Cherif}, \citenamefont {Stashkevich},
  \citenamefont {Jacques}, \citenamefont {Thiaville},\ and\ \citenamefont
  {Rohart}}]{RN24}%
  \BibitemOpen
  \bibfield  {author} {\bibinfo {author} {\bibfnamefont {A.}~\bibnamefont
  {Hrabec}}, \bibinfo {author} {\bibfnamefont {J.}~\bibnamefont {Sampaio}},
  \bibinfo {author} {\bibfnamefont {M.}~\bibnamefont {Belmeguenai}}, \bibinfo
  {author} {\bibfnamefont {I.}~\bibnamefont {Gross}}, \bibinfo {author}
  {\bibfnamefont {R.}~\bibnamefont {Weil}}, \bibinfo {author} {\bibfnamefont
  {S.~M.}\ \bibnamefont {Cherif}}, \bibinfo {author} {\bibfnamefont
  {A.}~\bibnamefont {Stashkevich}}, \bibinfo {author} {\bibfnamefont
  {V.}~\bibnamefont {Jacques}}, \bibinfo {author} {\bibfnamefont
  {A.}~\bibnamefont {Thiaville}}, \ and\ \bibinfo {author} {\bibfnamefont
  {S.}~\bibnamefont {Rohart}},\ }\href@noop {} {\bibfield  {journal} {\bibinfo
  {journal} {Nature Communications}\ }\textbf {\bibinfo {volume} {8}},\
  \bibinfo {pages} {15765} (\bibinfo {year} {2017})}\BibitemShut {NoStop}%
\bibitem [{\citenamefont {Sampaio}\ \emph {et~al.}(2013)\citenamefont
  {Sampaio}, \citenamefont {Cros}, \citenamefont {Rohart}, \citenamefont
  {Thiaville},\ and\ \citenamefont {Fert}}]{RN25}%
  \BibitemOpen
  \bibfield  {author} {\bibinfo {author} {\bibfnamefont {J.}~\bibnamefont
  {Sampaio}}, \bibinfo {author} {\bibfnamefont {V.}~\bibnamefont {Cros}},
  \bibinfo {author} {\bibfnamefont {S.}~\bibnamefont {Rohart}}, \bibinfo
  {author} {\bibfnamefont {A.}~\bibnamefont {Thiaville}}, \ and\ \bibinfo
  {author} {\bibfnamefont {A.}~\bibnamefont {Fert}},\ }\href@noop {} {\bibfield
   {journal} {\bibinfo  {journal} {Nature Nanotechnology}\ }\textbf {\bibinfo
  {volume} {8}},\ \bibinfo {pages} {839} (\bibinfo {year} {2013})}\BibitemShut
  {NoStop}%
\bibitem [{\citenamefont {Yu}\ \emph {et~al.}(2016)\citenamefont {Yu},
  \citenamefont {Upadhyaya}, \citenamefont {Li}, \citenamefont {Li},
  \citenamefont {Kim}, \citenamefont {Fan}, \citenamefont {Wong}, \citenamefont
  {Tserkovnyak}, \citenamefont {Amiri},\ and\ \citenamefont {Wang}}]{RN23}%
  \BibitemOpen
  \bibfield  {author} {\bibinfo {author} {\bibfnamefont {G.}~\bibnamefont
  {Yu}}, \bibinfo {author} {\bibfnamefont {P.}~\bibnamefont {Upadhyaya}},
  \bibinfo {author} {\bibfnamefont {X.}~\bibnamefont {Li}}, \bibinfo {author}
  {\bibfnamefont {W.}~\bibnamefont {Li}}, \bibinfo {author} {\bibfnamefont
  {S.~K.}\ \bibnamefont {Kim}}, \bibinfo {author} {\bibfnamefont
  {Y.}~\bibnamefont {Fan}}, \bibinfo {author} {\bibfnamefont {K.~L.}\
  \bibnamefont {Wong}}, \bibinfo {author} {\bibfnamefont {Y.}~\bibnamefont
  {Tserkovnyak}}, \bibinfo {author} {\bibfnamefont {P.~K.}\ \bibnamefont
  {Amiri}}, \ and\ \bibinfo {author} {\bibfnamefont {K.~L.}\ \bibnamefont
  {Wang}},\ }\href@noop {} {\bibfield  {journal} {\bibinfo  {journal} {Nano
  Letters}\ }\textbf {\bibinfo {volume} {16}},\ \bibinfo {pages} {1981}
  (\bibinfo {year} {2016})}\BibitemShut {NoStop}%
\bibitem [{\citenamefont {Buttner}\ \emph {et~al.}(2017)\citenamefont
  {Buttner}, \citenamefont {Lemesh}, \citenamefont {Schneider}, \citenamefont
  {Pfau}, \citenamefont {Gunther}, \citenamefont {Hessing}, \citenamefont
  {Geilhufe}, \citenamefont {Caretta}, \citenamefont {Engel}, \citenamefont
  {Kruger}, \citenamefont {Viefhaus}, \citenamefont {Eisebitt},\ and\
  \citenamefont {Beach}}]{RN29}%
  \BibitemOpen
  \bibfield  {author} {\bibinfo {author} {\bibfnamefont {F.}~\bibnamefont
  {Buttner}}, \bibinfo {author} {\bibfnamefont {I.}~\bibnamefont {Lemesh}},
  \bibinfo {author} {\bibfnamefont {M.}~\bibnamefont {Schneider}}, \bibinfo
  {author} {\bibfnamefont {B.}~\bibnamefont {Pfau}}, \bibinfo {author}
  {\bibfnamefont {C.~M.}\ \bibnamefont {Gunther}}, \bibinfo {author}
  {\bibfnamefont {P.}~\bibnamefont {Hessing}}, \bibinfo {author} {\bibfnamefont
  {J.}~\bibnamefont {Geilhufe}}, \bibinfo {author} {\bibfnamefont
  {L.}~\bibnamefont {Caretta}}, \bibinfo {author} {\bibfnamefont
  {D.}~\bibnamefont {Engel}}, \bibinfo {author} {\bibfnamefont
  {B.}~\bibnamefont {Kruger}}, \bibinfo {author} {\bibfnamefont
  {J.}~\bibnamefont {Viefhaus}}, \bibinfo {author} {\bibfnamefont
  {S.}~\bibnamefont {Eisebitt}}, \ and\ \bibinfo {author} {\bibfnamefont
  {G.~S.~D.}\ \bibnamefont {Beach}},\ }\href@noop {} {\bibfield  {journal}
  {\bibinfo  {journal} {Nature Nanotechnology}\ }\textbf {\bibinfo {volume}
  {12}},\ \bibinfo {pages} {1040} (\bibinfo {year} {2017})}\BibitemShut
  {NoStop}%
\bibitem [{\citenamefont {Wang}\ \emph
  {et~al.}(2018{\natexlab{a}})\citenamefont {Wang}, \citenamefont {Feng},
  \citenamefont {Kim}, \citenamefont {Kim}, \citenamefont {Lee}, \citenamefont
  {Pollard}, \citenamefont {Shin}, \citenamefont {Zhou}, \citenamefont {Peng},
  \citenamefont {Lee}, \citenamefont {Meng}, \citenamefont {Yang},
  \citenamefont {Han}, \citenamefont {Kim}, \citenamefont {Lu},\ and\
  \citenamefont {Noh}}]{Noh2018}%
  \BibitemOpen
  \bibfield  {author} {\bibinfo {author} {\bibfnamefont {L.}~\bibnamefont
  {Wang}}, \bibinfo {author} {\bibfnamefont {Q.}~\bibnamefont {Feng}}, \bibinfo
  {author} {\bibfnamefont {Y.}~\bibnamefont {Kim}}, \bibinfo {author}
  {\bibfnamefont {R.}~\bibnamefont {Kim}}, \bibinfo {author} {\bibfnamefont
  {K.~H.}\ \bibnamefont {Lee}}, \bibinfo {author} {\bibfnamefont {S.~D.}\
  \bibnamefont {Pollard}}, \bibinfo {author} {\bibfnamefont {Y.~J.}\
  \bibnamefont {Shin}}, \bibinfo {author} {\bibfnamefont {H.}~\bibnamefont
  {Zhou}}, \bibinfo {author} {\bibfnamefont {W.}~\bibnamefont {Peng}}, \bibinfo
  {author} {\bibfnamefont {D.}~\bibnamefont {Lee}}, \bibinfo {author}
  {\bibfnamefont {W.}~\bibnamefont {Meng}}, \bibinfo {author} {\bibfnamefont
  {H.}~\bibnamefont {Yang}}, \bibinfo {author} {\bibfnamefont {J.~H.}\
  \bibnamefont {Han}}, \bibinfo {author} {\bibfnamefont {M.}~\bibnamefont
  {Kim}}, \bibinfo {author} {\bibfnamefont {Q.}~\bibnamefont {Lu}}, \ and\
  \bibinfo {author} {\bibfnamefont {T.~W.}\ \bibnamefont {Noh}},\ }\href@noop
  {} {\bibfield  {journal} {\bibinfo  {journal} {Nature Materials}\ }\textbf
  {\bibinfo {volume} {17}},\ \bibinfo {pages} {1087} (\bibinfo {year}
  {2018}{\natexlab{a}})}\BibitemShut {NoStop}%
\bibitem [{\citenamefont {Liu}\ \emph {et~al.}(2017)\citenamefont {Liu},
  \citenamefont {Lei}, \citenamefont {Zhao}, \citenamefont {Liu}, \citenamefont
  {Ruotolo}, \citenamefont {Braun},\ and\ \citenamefont {Zhou}}]{993433}%
  \BibitemOpen
  \bibfield  {author} {\bibinfo {author} {\bibfnamefont {Y.}~\bibnamefont
  {Liu}}, \bibinfo {author} {\bibfnamefont {N.}~\bibnamefont {Lei}}, \bibinfo
  {author} {\bibfnamefont {W.}~\bibnamefont {Zhao}}, \bibinfo {author}
  {\bibfnamefont {W.}~\bibnamefont {Liu}}, \bibinfo {author} {\bibfnamefont
  {A.}~\bibnamefont {Ruotolo}}, \bibinfo {author} {\bibfnamefont {H.-B.}\
  \bibnamefont {Braun}}, \ and\ \bibinfo {author} {\bibfnamefont
  {Y.}~\bibnamefont {Zhou}},\ }\href@noop {} {\bibfield  {journal} {\bibinfo
  {journal} {Applied Physics Letters}\ }\textbf {\bibinfo {volume} {111}},\
  \bibinfo {pages} {022406} (\bibinfo {year} {2017})}\BibitemShut {NoStop}%
\bibitem [{\citenamefont {Wang}\ \emph
  {et~al.}(2018{\natexlab{b}})\citenamefont {Wang}, \citenamefont {Shi},\ and\
  \citenamefont {Kamlah}}]{PhysRevB.97.024429}%
  \BibitemOpen
  \bibfield  {author} {\bibinfo {author} {\bibfnamefont {J.}~\bibnamefont
  {Wang}}, \bibinfo {author} {\bibfnamefont {Y.}~\bibnamefont {Shi}}, \ and\
  \bibinfo {author} {\bibfnamefont {M.}~\bibnamefont {Kamlah}},\ }\href@noop {}
  {\bibfield  {journal} {\bibinfo  {journal} {Phys. Rev. B}\ }\textbf {\bibinfo
  {volume} {97}},\ \bibinfo {pages} {024429} (\bibinfo {year}
  {2018}{\natexlab{b}})}\BibitemShut {NoStop}%
\bibitem [{\citenamefont {Sadovnikov}\ \emph {et~al.}(2018)\citenamefont
  {Sadovnikov}, \citenamefont {Grachev}, \citenamefont {Sheshukova},
  \citenamefont {Sharaevskii}, \citenamefont {Serdobintsev}, \citenamefont
  {Mitin},\ and\ \citenamefont {Nikitov}}]{Nikitov2018}%
  \BibitemOpen
  \bibfield  {author} {\bibinfo {author} {\bibfnamefont {A.~V.}\ \bibnamefont
  {Sadovnikov}}, \bibinfo {author} {\bibfnamefont {A.~A.}\ \bibnamefont
  {Grachev}}, \bibinfo {author} {\bibfnamefont {S.~E.}\ \bibnamefont
  {Sheshukova}}, \bibinfo {author} {\bibfnamefont {Y.~P.}\ \bibnamefont
  {Sharaevskii}}, \bibinfo {author} {\bibfnamefont {A.~A.}\ \bibnamefont
  {Serdobintsev}}, \bibinfo {author} {\bibfnamefont {D.~M.}\ \bibnamefont
  {Mitin}}, \ and\ \bibinfo {author} {\bibfnamefont {S.~A.}\ \bibnamefont
  {Nikitov}},\ }\href@noop {} {\bibfield  {journal} {\bibinfo  {journal} {Phys.
  Rev. Lett.}\ }\textbf {\bibinfo {volume} {120}},\ \bibinfo {pages} {257203}
  (\bibinfo {year} {2018})}\BibitemShut {NoStop}%
\bibitem [{\citenamefont {Shi}\ and\ \citenamefont
  {Wang}(2018)}]{PhysRevB.97.224428}%
  \BibitemOpen
  \bibfield  {author} {\bibinfo {author} {\bibfnamefont {Y.}~\bibnamefont
  {Shi}}\ and\ \bibinfo {author} {\bibfnamefont {J.}~\bibnamefont {Wang}},\
  }\href@noop {} {\bibfield  {journal} {\bibinfo  {journal} {Phys. Rev. B}\
  }\textbf {\bibinfo {volume} {97}},\ \bibinfo {pages} {224428} (\bibinfo
  {year} {2018})}\BibitemShut {NoStop}%
\bibitem [{\citenamefont {Sun}\ \emph {et~al.}(2017)\citenamefont {Sun},
  \citenamefont {Ba}, \citenamefont {Chen}, \citenamefont {He}, \citenamefont
  {Wang}, \citenamefont {Zheng}, \citenamefont {Zou}, \citenamefont {Zhang},
  \citenamefont {Yang}, \citenamefont {Yan}, \citenamefont {Feng},
  \citenamefont {Zhang}, \citenamefont {Cai}, \citenamefont {Wu}, \citenamefont
  {Liu}, \citenamefont {Gu}, \citenamefont {Cheng}, \citenamefont {Nan},
  \citenamefont {Qiu}, \citenamefont {Wu}, \citenamefont {Li},\ and\
  \citenamefont {Zhao}}]{RN31}%
  \BibitemOpen
  \bibfield  {author} {\bibinfo {author} {\bibfnamefont {Y.}~\bibnamefont
  {Sun}}, \bibinfo {author} {\bibfnamefont {Y.}~\bibnamefont {Ba}}, \bibinfo
  {author} {\bibfnamefont {A.}~\bibnamefont {Chen}}, \bibinfo {author}
  {\bibfnamefont {W.}~\bibnamefont {He}}, \bibinfo {author} {\bibfnamefont
  {W.}~\bibnamefont {Wang}}, \bibinfo {author} {\bibfnamefont {X.}~\bibnamefont
  {Zheng}}, \bibinfo {author} {\bibfnamefont {L.}~\bibnamefont {Zou}}, \bibinfo
  {author} {\bibfnamefont {Y.}~\bibnamefont {Zhang}}, \bibinfo {author}
  {\bibfnamefont {Q.}~\bibnamefont {Yang}}, \bibinfo {author} {\bibfnamefont
  {L.}~\bibnamefont {Yan}}, \bibinfo {author} {\bibfnamefont {C.}~\bibnamefont
  {Feng}}, \bibinfo {author} {\bibfnamefont {Q.}~\bibnamefont {Zhang}},
  \bibinfo {author} {\bibfnamefont {J.}~\bibnamefont {Cai}}, \bibinfo {author}
  {\bibfnamefont {W.}~\bibnamefont {Wu}}, \bibinfo {author} {\bibfnamefont
  {M.}~\bibnamefont {Liu}}, \bibinfo {author} {\bibfnamefont {L.}~\bibnamefont
  {Gu}}, \bibinfo {author} {\bibfnamefont {Z.}~\bibnamefont {Cheng}}, \bibinfo
  {author} {\bibfnamefont {C.-W.}\ \bibnamefont {Nan}}, \bibinfo {author}
  {\bibfnamefont {Z.}~\bibnamefont {Qiu}}, \bibinfo {author} {\bibfnamefont
  {Y.}~\bibnamefont {Wu}}, \bibinfo {author} {\bibfnamefont {J.}~\bibnamefont
  {Li}}, \ and\ \bibinfo {author} {\bibfnamefont {Y.}~\bibnamefont {Zhao}},\
  }\href@noop {} {\bibfield  {journal} {\bibinfo  {journal} {ACS Applied
  Materials and Interfaces}\ }\textbf {\bibinfo {volume} {9}},\ \bibinfo
  {pages} {10855} (\bibinfo {year} {2017})}\BibitemShut {NoStop}%
\bibitem [{\citenamefont {Dieny}\ and\ \citenamefont
  {Chshiev}(2017)}]{RevModPhys.89.025008}%
  \BibitemOpen
  \bibfield  {author} {\bibinfo {author} {\bibfnamefont {B.}~\bibnamefont
  {Dieny}}\ and\ \bibinfo {author} {\bibfnamefont {M.}~\bibnamefont
  {Chshiev}},\ }\href@noop {} {\bibfield  {journal} {\bibinfo  {journal} {Rev.
  Mod. Phys.}\ }\textbf {\bibinfo {volume} {89}},\ \bibinfo {pages} {025008}
  (\bibinfo {year} {2017})}\BibitemShut {NoStop}%
\bibitem [{\citenamefont {Nii}\ \emph {et~al.}(2015)\citenamefont {Nii},
  \citenamefont {Nakajima}, \citenamefont {Kikkawa}, \citenamefont {Yamasaki},
  \citenamefont {Ohishi}, \citenamefont {Suzuki}, \citenamefont {Taguchi},
  \citenamefont {Arima}, \citenamefont {Tokura},\ and\ \citenamefont
  {Iwasa}}]{Iwasa2015}%
  \BibitemOpen
  \bibfield  {author} {\bibinfo {author} {\bibfnamefont {Y.}~\bibnamefont
  {Nii}}, \bibinfo {author} {\bibfnamefont {T.}~\bibnamefont {Nakajima}},
  \bibinfo {author} {\bibfnamefont {A.}~\bibnamefont {Kikkawa}}, \bibinfo
  {author} {\bibfnamefont {Y.}~\bibnamefont {Yamasaki}}, \bibinfo {author}
  {\bibfnamefont {K.}~\bibnamefont {Ohishi}}, \bibinfo {author} {\bibfnamefont
  {J.}~\bibnamefont {Suzuki}}, \bibinfo {author} {\bibfnamefont
  {Y.}~\bibnamefont {Taguchi}}, \bibinfo {author} {\bibfnamefont
  {T.}~\bibnamefont {Arima}}, \bibinfo {author} {\bibfnamefont
  {Y.}~\bibnamefont {Tokura}}, \ and\ \bibinfo {author} {\bibfnamefont
  {Y.}~\bibnamefont {Iwasa}},\ }\href@noop {} {\bibfield  {journal} {\bibinfo
  {journal} {Nature Communication}\ }\textbf {\bibinfo {volume} {6}},\ \bibinfo
  {pages} {8539} (\bibinfo {year} {2015})}\BibitemShut {NoStop}%
\bibitem [{\citenamefont {Shibata}\ \emph {et~al.}(2015)\citenamefont
  {Shibata}, \citenamefont {Iwasaki}, \citenamefont {Kanazawa}, \citenamefont
  {Aizawa}, \citenamefont {Tanigaki}, \citenamefont {Shirai}, \citenamefont
  {Nakajima}, \citenamefont {Kubota}, \citenamefont {Kawasaki}, \citenamefont
  {Park}, \citenamefont {Shindo}, \citenamefont {Nagaosa},\ and\ \citenamefont
  {Tokura}}]{RN20}%
  \BibitemOpen
  \bibfield  {author} {\bibinfo {author} {\bibfnamefont {K.}~\bibnamefont
  {Shibata}}, \bibinfo {author} {\bibfnamefont {J.}~\bibnamefont {Iwasaki}},
  \bibinfo {author} {\bibfnamefont {N.}~\bibnamefont {Kanazawa}}, \bibinfo
  {author} {\bibfnamefont {S.}~\bibnamefont {Aizawa}}, \bibinfo {author}
  {\bibfnamefont {T.}~\bibnamefont {Tanigaki}}, \bibinfo {author}
  {\bibfnamefont {M.}~\bibnamefont {Shirai}}, \bibinfo {author} {\bibfnamefont
  {T.}~\bibnamefont {Nakajima}}, \bibinfo {author} {\bibfnamefont
  {M.}~\bibnamefont {Kubota}}, \bibinfo {author} {\bibfnamefont
  {M.}~\bibnamefont {Kawasaki}}, \bibinfo {author} {\bibfnamefont {H.~S.}\
  \bibnamefont {Park}}, \bibinfo {author} {\bibfnamefont {D.}~\bibnamefont
  {Shindo}}, \bibinfo {author} {\bibfnamefont {N.}~\bibnamefont {Nagaosa}}, \
  and\ \bibinfo {author} {\bibfnamefont {Y.}~\bibnamefont {Tokura}},\
  }\href@noop {} {\bibfield  {journal} {\bibinfo  {journal} {Nature
  Nanotechnology}\ }\textbf {\bibinfo {volume} {10}},\ \bibinfo {pages} {589}
  (\bibinfo {year} {2015})}\BibitemShut {NoStop}%
\bibitem [{\citenamefont {Koretsune}\ \emph {et~al.}(2015)\citenamefont
  {Koretsune}, \citenamefont {Nagaosa},\ and\ \citenamefont
  {Arita}}]{Arita2015}%
  \BibitemOpen
  \bibfield  {author} {\bibinfo {author} {\bibfnamefont {T.}~\bibnamefont
  {Koretsune}}, \bibinfo {author} {\bibfnamefont {N.}~\bibnamefont {Nagaosa}},
  \ and\ \bibinfo {author} {\bibfnamefont {R.}~\bibnamefont {Arita}},\ }\href
  {\doibase 10.1038/srep13302} {\bibfield  {journal} {\bibinfo  {journal}
  {Scientific Reports}\ }\textbf {\bibinfo {volume} {5}},\ \bibinfo {pages}
  {13302} (\bibinfo {year} {2015})}\BibitemShut {NoStop}%
\bibitem [{\citenamefont {Baraduc}\ \emph {et~al.}(2018)\citenamefont
  {Baraduc}, \citenamefont {Srivastava}, \citenamefont {Schott}, \citenamefont
  {Belmeguenai}, \citenamefont {Roussigné}, \citenamefont {Bernand-Mantel},
  \citenamefont {Ranno}, \citenamefont {Pizzini}, \citenamefont {Chérif},
  \citenamefont {Stashkevich}, \citenamefont {Auffret}, \citenamefont
  {Chshiev},\ and\ \citenamefont {Béa}}]{Bea2018}%
  \BibitemOpen
  \bibfield  {author} {\bibinfo {author} {\bibfnamefont {C.}~\bibnamefont
  {Baraduc}}, \bibinfo {author} {\bibfnamefont {T.}~\bibnamefont {Srivastava}},
  \bibinfo {author} {\bibfnamefont {M.}~\bibnamefont {Schott}}, \bibinfo
  {author} {\bibfnamefont {M.}~\bibnamefont {Belmeguenai}}, \bibinfo {author}
  {\bibfnamefont {Y.}~\bibnamefont {Roussigné}}, \bibinfo {author}
  {\bibfnamefont {A.}~\bibnamefont {Bernand-Mantel}}, \bibinfo {author}
  {\bibfnamefont {L.}~\bibnamefont {Ranno}}, \bibinfo {author} {\bibfnamefont
  {S.}~\bibnamefont {Pizzini}}, \bibinfo {author} {\bibfnamefont {S.~M.}\
  \bibnamefont {Chérif}}, \bibinfo {author} {\bibfnamefont {A.}~\bibnamefont
  {Stashkevich}}, \bibinfo {author} {\bibfnamefont {S.}~\bibnamefont
  {Auffret}}, \bibinfo {author} {\bibfnamefont {M.}~\bibnamefont {Chshiev}}, \
  and\ \bibinfo {author} {\bibfnamefont {H.}~\bibnamefont {Béa}},\ }\href@noop
  {} {\bibfield  {journal} {\bibinfo  {journal} {Proceedings, Spintronics XI}\
  }\textbf {\bibinfo {volume} {10732}} (\bibinfo {year} {2018})}\BibitemShut
  {NoStop}%
\bibitem [{\citenamefont {Davydenko}\ \emph {et~al.}(2019)\citenamefont
  {Davydenko}, \citenamefont {Kozlov}, \citenamefont {Kolesnikov},
  \citenamefont {Stebliy}, \citenamefont {Suslin}, \citenamefont {Vekovshinin},
  \citenamefont {Sadovnikov},\ and\ \citenamefont
  {Nikitov}}]{PhysRevB.99.014433}%
  \BibitemOpen
  \bibfield  {author} {\bibinfo {author} {\bibfnamefont {A.~V.}\ \bibnamefont
  {Davydenko}}, \bibinfo {author} {\bibfnamefont {A.~G.}\ \bibnamefont
  {Kozlov}}, \bibinfo {author} {\bibfnamefont {A.~G.}\ \bibnamefont
  {Kolesnikov}}, \bibinfo {author} {\bibfnamefont {M.~E.}\ \bibnamefont
  {Stebliy}}, \bibinfo {author} {\bibfnamefont {G.~S.}\ \bibnamefont {Suslin}},
  \bibinfo {author} {\bibfnamefont {Y.~E.}\ \bibnamefont {Vekovshinin}},
  \bibinfo {author} {\bibfnamefont {A.~V.}\ \bibnamefont {Sadovnikov}}, \ and\
  \bibinfo {author} {\bibfnamefont {S.~A.}\ \bibnamefont {Nikitov}},\ }\href
  {\doibase 10.1103/PhysRevB.99.014433} {\bibfield  {journal} {\bibinfo
  {journal} {Phys. Rev. B}\ }\textbf {\bibinfo {volume} {99}},\ \bibinfo
  {pages} {014433} (\bibinfo {year} {2019})}\BibitemShut {NoStop}%
\bibitem [{\citenamefont {Tacchi}\ \emph {et~al.}(2017)\citenamefont {Tacchi},
  \citenamefont {Troncoso}, \citenamefont {Ahlberg}, \citenamefont {Gubbiotti},
  \citenamefont {Madami}, \citenamefont {\AA{}kerman},\ and\ \citenamefont
  {Landeros}}]{PhysRevLett.118.147201}%
  \BibitemOpen
  \bibfield  {author} {\bibinfo {author} {\bibfnamefont {S.}~\bibnamefont
  {Tacchi}}, \bibinfo {author} {\bibfnamefont {R.~E.}\ \bibnamefont
  {Troncoso}}, \bibinfo {author} {\bibfnamefont {M.}~\bibnamefont {Ahlberg}},
  \bibinfo {author} {\bibfnamefont {G.}~\bibnamefont {Gubbiotti}}, \bibinfo
  {author} {\bibfnamefont {M.}~\bibnamefont {Madami}}, \bibinfo {author}
  {\bibfnamefont {J.}~\bibnamefont {\AA{}kerman}}, \ and\ \bibinfo {author}
  {\bibfnamefont {P.}~\bibnamefont {Landeros}},\ }\href {\doibase
  10.1103/PhysRevLett.118.147201} {\bibfield  {journal} {\bibinfo  {journal}
  {Phys. Rev. Lett.}\ }\textbf {\bibinfo {volume} {118}},\ \bibinfo {pages}
  {147201} (\bibinfo {year} {2017})}\BibitemShut {NoStop}%
\bibitem [{\citenamefont {Landau}\ and\ \citenamefont
  {Lifshitz}(1970)}]{LL_elast}%
  \BibitemOpen
  \bibfield  {author} {\bibinfo {author} {\bibfnamefont {L.}~\bibnamefont
  {Landau}}\ and\ \bibinfo {author} {\bibfnamefont {E.}~\bibnamefont
  {Lifshitz}},\ }\href@noop {} {\emph {\bibinfo {title} {Theory of Elasticity (
  Volume 7 of A Course of Theoretical Physics )}}}\ (\bibinfo  {publisher}
  {Pergamon Press},\ \bibinfo {year} {1970})\BibitemShut {NoStop}%
\bibitem [{\citenamefont {Di}\ \emph {et~al.}(2015)\citenamefont {Di},
  \citenamefont {Zhang}, \citenamefont {Lim}, \citenamefont {Ng}, \citenamefont
  {Kuok}, \citenamefont {Yu}, \citenamefont {Yoon}, \citenamefont {Qiu},\ and\
  \citenamefont {Yang}}]{PhysRevLett.114.047201}%
  \BibitemOpen
  \bibfield  {author} {\bibinfo {author} {\bibfnamefont {K.}~\bibnamefont
  {Di}}, \bibinfo {author} {\bibfnamefont {V.~L.}\ \bibnamefont {Zhang}},
  \bibinfo {author} {\bibfnamefont {H.~S.}\ \bibnamefont {Lim}}, \bibinfo
  {author} {\bibfnamefont {S.~C.}\ \bibnamefont {Ng}}, \bibinfo {author}
  {\bibfnamefont {M.~H.}\ \bibnamefont {Kuok}}, \bibinfo {author}
  {\bibfnamefont {J.}~\bibnamefont {Yu}}, \bibinfo {author} {\bibfnamefont
  {J.}~\bibnamefont {Yoon}}, \bibinfo {author} {\bibfnamefont {X.}~\bibnamefont
  {Qiu}}, \ and\ \bibinfo {author} {\bibfnamefont {H.}~\bibnamefont {Yang}},\
  }\href {\doibase 10.1103/PhysRevLett.114.047201} {\bibfield  {journal}
  {\bibinfo  {journal} {Phys. Rev. Lett.}\ }\textbf {\bibinfo {volume} {114}},\
  \bibinfo {pages} {047201} (\bibinfo {year} {2015})}\BibitemShut {NoStop}%
\bibitem [{\citenamefont {Moon}\ \emph {et~al.}(2013)\citenamefont {Moon},
  \citenamefont {Seo}, \citenamefont {Lee}, \citenamefont {Kim}, \citenamefont
  {Ryu}, \citenamefont {Lee}, \citenamefont {McMichael},\ and\ \citenamefont
  {Stiles}}]{PhysRevB.88.184404}%
  \BibitemOpen
  \bibfield  {author} {\bibinfo {author} {\bibfnamefont {J.-H.}\ \bibnamefont
  {Moon}}, \bibinfo {author} {\bibfnamefont {S.-M.}\ \bibnamefont {Seo}},
  \bibinfo {author} {\bibfnamefont {K.-J.}\ \bibnamefont {Lee}}, \bibinfo
  {author} {\bibfnamefont {K.-W.}\ \bibnamefont {Kim}}, \bibinfo {author}
  {\bibfnamefont {J.}~\bibnamefont {Ryu}}, \bibinfo {author} {\bibfnamefont
  {H.-W.}\ \bibnamefont {Lee}}, \bibinfo {author} {\bibfnamefont {R.~D.}\
  \bibnamefont {McMichael}}, \ and\ \bibinfo {author} {\bibfnamefont {M.~D.}\
  \bibnamefont {Stiles}},\ }\href {\doibase 10.1103/PhysRevB.88.184404}
  {\bibfield  {journal} {\bibinfo  {journal} {Phys. Rev. B}\ }\textbf {\bibinfo
  {volume} {88}},\ \bibinfo {pages} {184404} (\bibinfo {year}
  {2013})}\BibitemShut {NoStop}%
\bibitem [{\citenamefont {Sun}\ \emph {et~al.}(2013)\citenamefont {Sun},
  \citenamefont {Cao}, \citenamefont {Miao}, \citenamefont {Feng},
  \citenamefont {You}, \citenamefont {Wu}, \citenamefont {Zhang}, \citenamefont
  {Hu},\ and\ \citenamefont {Ding}}]{Ding2013}%
  \BibitemOpen
  \bibfield  {author} {\bibinfo {author} {\bibfnamefont {L.}~\bibnamefont
  {Sun}}, \bibinfo {author} {\bibfnamefont {R.~X.}\ \bibnamefont {Cao}},
  \bibinfo {author} {\bibfnamefont {B.~F.}\ \bibnamefont {Miao}}, \bibinfo
  {author} {\bibfnamefont {Z.}~\bibnamefont {Feng}}, \bibinfo {author}
  {\bibfnamefont {B.}~\bibnamefont {You}}, \bibinfo {author} {\bibfnamefont
  {D.}~\bibnamefont {Wu}}, \bibinfo {author} {\bibfnamefont {W.}~\bibnamefont
  {Zhang}}, \bibinfo {author} {\bibfnamefont {A.}~\bibnamefont {Hu}}, \ and\
  \bibinfo {author} {\bibfnamefont {H.~F.}\ \bibnamefont {Ding}},\ }\href@noop
  {} {\bibfield  {journal} {\bibinfo  {journal} {Phys. Rev. Lett.}\ }\textbf
  {\bibinfo {volume} {110}},\ \bibinfo {pages} {167201} (\bibinfo {year}
  {2013})}\BibitemShut {NoStop}%
\bibitem [{\citenamefont {Eyrich}\ \emph {et~al.}(2012)\citenamefont {Eyrich},
  \citenamefont {Huttema}, \citenamefont {Arora}, \citenamefont {Montoya},
  \citenamefont {Rashidi}, \citenamefont {Burrowes}, \citenamefont {Kardasz},
  \citenamefont {Girt}, \citenamefont {Heinrich}, \citenamefont {Mryasov},
  \citenamefont {From},\ and\ \citenamefont {Karis}}]{Karis2012}%
  \BibitemOpen
  \bibfield  {author} {\bibinfo {author} {\bibfnamefont {C.}~\bibnamefont
  {Eyrich}}, \bibinfo {author} {\bibfnamefont {W.}~\bibnamefont {Huttema}},
  \bibinfo {author} {\bibfnamefont {M.}~\bibnamefont {Arora}}, \bibinfo
  {author} {\bibfnamefont {E.}~\bibnamefont {Montoya}}, \bibinfo {author}
  {\bibfnamefont {F.}~\bibnamefont {Rashidi}}, \bibinfo {author} {\bibfnamefont
  {C.}~\bibnamefont {Burrowes}}, \bibinfo {author} {\bibfnamefont
  {B.}~\bibnamefont {Kardasz}}, \bibinfo {author} {\bibfnamefont
  {E.}~\bibnamefont {Girt}}, \bibinfo {author} {\bibfnamefont {B.}~\bibnamefont
  {Heinrich}}, \bibinfo {author} {\bibfnamefont {O.~N.}\ \bibnamefont
  {Mryasov}}, \bibinfo {author} {\bibfnamefont {M.}~\bibnamefont {From}}, \
  and\ \bibinfo {author} {\bibfnamefont {O.}~\bibnamefont {Karis}},\
  }\href@noop {} {\bibfield  {journal} {\bibinfo  {journal} {J. Appl. Phys.}\
  }\textbf {\bibinfo {volume} {111}},\ \bibinfo {pages} {07C919} (\bibinfo
  {year} {2012})}\BibitemShut {NoStop}%
\bibitem [{\citenamefont {Fert}\ and\ \citenamefont
  {Levy}(1980)}]{PhysRevLett.44.1538}%
  \BibitemOpen
  \bibfield  {author} {\bibinfo {author} {\bibfnamefont {A.}~\bibnamefont
  {Fert}}\ and\ \bibinfo {author} {\bibfnamefont {P.~M.}\ \bibnamefont
  {Levy}},\ }\href {\doibase 10.1103/PhysRevLett.44.1538} {\bibfield  {journal}
  {\bibinfo  {journal} {Phys. Rev. Lett.}\ }\textbf {\bibinfo {volume} {44}},\
  \bibinfo {pages} {1538} (\bibinfo {year} {1980})}\BibitemShut {NoStop}%
\bibitem [{\citenamefont {Camosi}\ \emph {et~al.}(2017)\citenamefont {Camosi},
  \citenamefont {Rohart}, \citenamefont {Fruchart}, \citenamefont {Pizzini},
  \citenamefont {Belmeguenai}, \citenamefont {Roussigne}, \citenamefont
  {Stashkevich}, \citenamefont {Cherif}, \citenamefont {Ranno}, \citenamefont
  {de~Santis},\ and\ \citenamefont {Vogel}}]{PhysRevB.95.214422}%
  \BibitemOpen
  \bibfield  {author} {\bibinfo {author} {\bibfnamefont {L.}~\bibnamefont
  {Camosi}}, \bibinfo {author} {\bibfnamefont {S.}~\bibnamefont {Rohart}},
  \bibinfo {author} {\bibfnamefont {O.}~\bibnamefont {Fruchart}}, \bibinfo
  {author} {\bibfnamefont {S.}~\bibnamefont {Pizzini}}, \bibinfo {author}
  {\bibfnamefont {M.}~\bibnamefont {Belmeguenai}}, \bibinfo {author}
  {\bibfnamefont {Y.}~\bibnamefont {Roussigne}}, \bibinfo {author}
  {\bibfnamefont {A.}~\bibnamefont {Stashkevich}}, \bibinfo {author}
  {\bibfnamefont {S.~M.}\ \bibnamefont {Cherif}}, \bibinfo {author}
  {\bibfnamefont {L.}~\bibnamefont {Ranno}}, \bibinfo {author} {\bibfnamefont
  {M.}~\bibnamefont {de~Santis}}, \ and\ \bibinfo {author} {\bibfnamefont
  {J.}~\bibnamefont {Vogel}},\ }\href@noop {} {\bibfield  {journal} {\bibinfo
  {journal} {Phys. Rev. B}\ }\textbf {\bibinfo {volume} {95}},\ \bibinfo
  {pages} {214422} (\bibinfo {year} {2017})}\BibitemShut {NoStop}%
\bibitem [{\citenamefont {Sohn}\ \emph {et~al.}(2015)\citenamefont {Sohn},
  \citenamefont {Nowakowski}, \citenamefont {yen Liang}, \citenamefont
  {Hockel}, \citenamefont {Wetzlar}, \citenamefont {Keller}, \citenamefont
  {McLellan}, \citenamefont {Marcus}, \citenamefont {Doran}, \citenamefont
  {Young}, \citenamefont {Klaui}, \citenamefont {Carman},\ and\ \citenamefont
  {Jeffrey~Bokor}}]{Carman}%
  \BibitemOpen
  \bibfield  {author} {\bibinfo {author} {\bibfnamefont {H.}~\bibnamefont
  {Sohn}}, \bibinfo {author} {\bibfnamefont {M.~E.}\ \bibnamefont
  {Nowakowski}}, \bibinfo {author} {\bibfnamefont {C.}~\bibnamefont {yen
  Liang}}, \bibinfo {author} {\bibfnamefont {J.~L.}\ \bibnamefont {Hockel}},
  \bibinfo {author} {\bibfnamefont {K.}~\bibnamefont {Wetzlar}}, \bibinfo
  {author} {\bibfnamefont {S.}~\bibnamefont {Keller}}, \bibinfo {author}
  {\bibfnamefont {B.~M.}\ \bibnamefont {McLellan}}, \bibinfo {author}
  {\bibfnamefont {M.~A.}\ \bibnamefont {Marcus}}, \bibinfo {author}
  {\bibfnamefont {A.}~\bibnamefont {Doran}}, \bibinfo {author} {\bibfnamefont
  {A.}~\bibnamefont {Young}}, \bibinfo {author} {\bibfnamefont
  {M.}~\bibnamefont {Klaui}}, \bibinfo {author} {\bibfnamefont {G.~P.}\
  \bibnamefont {Carman}}, \ and\ \bibinfo {author} {\bibfnamefont {a.~R.
  N.~C.}\ \bibnamefont {Jeffrey~Bokor}},\ }\href@noop {} {\bibfield  {journal}
  {\bibinfo  {journal} {ASC Nano}\ }\textbf {\bibinfo {volume} {9}},\ \bibinfo
  {pages} {4814} (\bibinfo {year} {2015})}\BibitemShut {NoStop}%
\bibitem [{\citenamefont {Donahue}\ and\ \citenamefont {Porter}(1999)}]{OOMMF}%
  \BibitemOpen
  \bibfield  {author} {\bibinfo {author} {\bibfnamefont {M.}~\bibnamefont
  {Donahue}}\ and\ \bibinfo {author} {\bibfnamefont {D.}~\bibnamefont
  {Porter}},\ }\href {\doibase 10.6028/NIST.IR.6376} {\emph {\bibinfo {title}
  {OOMMF User's Guide Version 1.0}}}\ (\bibinfo  {publisher} {National
  Institute of Standards and Technology, Gaithersburg, MDs},\ \bibinfo {year}
  {1999})\BibitemShut {NoStop}%
\end{thebibliography}%
\end{document}